\documentclass[longauth]{aa}  

\usepackage{array}
\usepackage{graphicx}
\usepackage{multirow}
\usepackage{color}
\usepackage{hyperref}
\usepackage{makecell}
\usepackage{longtable}
\usepackage[capitalize]{cleveref}
\usepackage{txfonts}
\renewcommand{\arraystretch}{1.2}

\begin{document}

   \title{A relativistic inflow candidate in a quasar at cosmic noon}

   % \subtitle{TBD}
   \titlerunning{A relativistic inflow candidate in a high-redshift quasar}
   \authorrunning{Y. Xu et al.}
   
   \author{Yerong Xu\inst{1,2}\corrauth{yerong.xu@ice.csic.es},
            Xiurui Zhao\inst{3}  \email{xiuruiz@caltech.edu},
            Elena Bertola\inst{4} \email{elena.bertola@inaf.it},
            Valentina Braito\inst{5,6,7} \email{valentina.braito@inaf.it},
            Francesca Civano\inst{8} \email{francesca.m.civano@nasa.gov},
            Luca Comisso\inst{9} \email{luca.comisso@columbia.edu},
            Luigi Gallo\inst{10} \email{luigi.gallo@smu.ca},
            Vittoria E. Gianolli\inst{11,6} \email{vgianol@clemson.edu},
            Elias Kammoun\inst{3} \email{ekammoun@caltech.edu},
            Giorgio Lanzuisi\inst{12} \email{giorgio.lanzuisi@inaf.it},
            Stefano Marchesi\inst{13,11,12} \email{stefano.marchesi@inaf.it},
            Mar Mezcua\inst{1,2} \email{mezcua@ice.csic.es},
            Ciro Pinto\inst{14} \email{ciro.pinto@inaf.it},
            James N. Reeves\inst{5,6} \email{reevesjn@cua.edu},
            Daniele Rogantini\inst{15} \email{danieler@uchicago.edu},
            N\'uria Torres-Alb\`a\inst{16} \email{ntorresalba@gmail.com},
            Lizhong Zhang\inst{9} \email{lizhong@ias.edu}
        }
        
   \institute{Institute of Space Sciences (ICE, CSIC), Campus UAB, Carrer de Magrans, 08193 Barcelona, Spain
   \and Institut d'Estudis Espacials de Catalunya (IEEC), Edifici RDIT, Campus UPC, 08860 Castelldefels, Barcelona, Spain
   \and Cahill Center for Astrophysics, California Institute of Technology, 1216 East California Boulevard, Pasadena, CA 91125, USA
   \and INAF-OAA, Osservatorio Astrofisico di Arcetri, largo E. Fermi 5, 50127, Firenze, Italy
   \and Institute for Astrophysics and Computational Sciences, Department of Physics, The Catholic University of America, Washington, DC 20064, USA
   \and INAF-Osservatorio Astronomico di Brera, Via Brera 28, 20121 Milano, Italy
   \and Dipartimento di Fisica, Universit\`a di Trento, Via Sommarive 14, Trento 38123, Italy
   \and NASA Goddard Space Flight Center, Greenbelt, MD 20771, USA
   \and Department of Astronomy and Columbia Astrophysics Laboratory, Columbia University, New York, NY 10027, USA
   \and Department of Astronomy \& Physics, Saint Mary’s University, 923 Robie Street, Halifax, NS B3H 3C3, Canada
   \and Department of Physics and Astronomy, Clemson University, Kinard Lab of Physics, Clemson, SC 29634, USA
   \and INAF - Osservatorio di Astrofisica e Scienza dello Spazio di Bologna, Via Piero Gobetti, 93/3, 40129, Bologna, Italy
   \and Dipartimento di Fisica e Astronomia ``Augusto Righi'', Università di Bologna, via Gobetti 93/2, I-40129 Bologna, Italy
   \and INAF – IASF Palermo, Via U. La Malfa 153, 90146 Palermo, Italy
   \and Department of Astronomy and Astrophysics, University of Chicago, Chicago, IL60637, USA
   \and Department of Physics, University of Virginia, Charlottesville, VA 22904, USA
             }

   \date{Received 02 October 2026 / Accepted 24 November 2026}

% \abstract{}{}{}{}{} 
% 5 {} token are mandatory

  \abstract
  % TBD
   {
   Cosmic noon ($z\sim2$) marks the peak epoch of cosmic star formation and black hole accretion activity, where strong accretion and outflows are expected. We report two candidate absorption features in the \textit{XMM-Newton} spectrum of the quasar WISEA~J094835.95+432302.6 at $z=1.893$. Two redshifted features occur at rest-frame energies of $\sim4.6$ and $\sim5.7$\,keV with confidence levels of 98.1\% and 88.8\%, respectively, as estimated from Monte Carlo simulations considering the look-elsewhere effect. The stronger feature is consistent with Fe XXV/XXVI K$\alpha$ absorption from an ultrafast inflow (UFI) candidate with an apparent velocity of $0.37^{+0.02}_{-0.01}c$, while higher-order Fe K transitions from the same absorber can partially contribute to the weaker feature. 
   If confirmed by follow-up \textit{XMM-Newton} observations, these detections would extend reported UFI candidates from the local Universe to cosmic noon, offering a rare opportunity to probe gas dynamics around a rapidly growing black hole at cosmic noon.
   }
  % % aims heading (mandatory)
   % {TBD}
  % % methods heading (mandatory)
  %  {TBD
  %  }
  % % results heading (mandatory)
  %  {TBD
  %  }
  % % conclusions heading (optional), leave it empty if necessary 
  %  {
  %  TBD
  %  }

   \keywords{accretion, accretion disks - quasars: individual: WISEA J094835.95+432302.6 -  quasars: absorption lines
                    }

   \maketitle
   \nolinenumbers
%
%-------------------------------------------------------------------

\section{Introduction}\label{sec:intro}
At cosmic noon ($z\sim2$), when both cosmic star formation and active galactic nucleus (AGN) activity peak, rapidly accreting supermassive black holes (SMBHs) are expected to undergo highly dynamic feeding and ejection, with inflowing gas coexisting with powerful winds or jets that can influence the host-galaxy interstellar medium \citep[ISM; e.g.,][]{2015King}. Probing accretion inflows and outflows at this epoch therefore provides a direct view of how SMBHs grow during their most active phase.
Over the past two decades, X-ray spectroscopy of AGNs have often revealed blueshifted absorption features from ionized winds. The most extreme cases are ultrafast outflows \citep[UFOs;][]{2010Tombesi}, with high ionization states ($\log\xi\sim3\mbox{--}6$), high column densities ($N_\mathrm{H}\sim10^{22}\mbox{--}10^{24}$\,cm$^{-2}$), and mildly relativistic velocities \citep[$v\sim0.03\mbox{--}0.5c$; e.g.,][]{2024Gianolli}. Their extreme velocities and variability point to an origin in the inner accretion disk. Their kinetic power can also represent a substantial fraction of the accretion budget, making UFOs a plausible channel through which the SMBH can communicate with its host galaxy \citep[e.g.,][]{2021Laha,2023Gallo}. However, UFO detection in high-redshift non-lensed systems requires deep exposures and is restricted to strong features \citep[$\mathrm{EW}\gtrsim50\mbox{--}250$\,eV in the $7\mbox{--}12$\,keV band;][]{2026Lanzuisi}.

Another relevant but much rarer phenomenon is significantly redshifted absorption features. They are often interpreted as highly ionized gas moving toward the SMBH along the line of sight (LOS), referred to as ultrafast inflows (UFIs). Only a few nearby AGNs have reported UFI candidates ($z\lesssim0.2$), including Mrk 509 \citep[$v_{\rm in}\sim0.15\mbox{--}0.20c$;][]{Dadina2005,Dadina2026}, Mrk 335 \citep[$v_{\rm in}\sim0.11\mbox{--}0.15c$;][]{Longinotti2007}, PG 1211+143 \citep[$v_{\rm in}\sim0.04\mbox{--}0.4c$;][]{Reeves2005,2018Pounds,2024Pounds}, E 1821+643 \citep[$v_\mathrm{in}\sim0.07\mbox{--}0.10c$;][]{Yaqoob2005}, NGC 2617 \citep[$v_{\rm in}\sim0.1c$;][]{Giustini2017}, and ESP 39607 \citep[$v_{\rm in}\sim0.15\mbox{--}0.3c$;][]{Peca2025,2026Peca}. Proposed origins include chaotic cold accretion \citep[e.g.,][]{Gaspari2013}, disk tearing in a misaligned accretion flow \citep{Dogan2018}, fallback from failed disk winds or aborted jets \citep{2004Proga,Ghisellini2004}, and magnetically channeled non-equatorial accretion \citep{Fukumura2026}. Their origin remains poorly constrained because of few known cases. In addition, UFIs may also provide insight into SMBH feeding and coronal variability. In Mrk 509, for example, they may seed inward-propagating fluctuations and contribute to the coronal variability (Zhao et al. in prep.).

In this letter, we report a UFI candidate in the quasar WISEA~J094835.95+432302.6 (hereafter J0948) at cosmic noon ($z=1.893$) observed by the Probing the AGN Coronae with High-redshift AGN (PACHA) project, which targets the X-ray coronae of luminous AGNs at $z>1$ using quasi-simultaneous \textit{XMM-Newton} and \textit{NuSTAR} data \citep{Zhao2026}. J0948 has an intrinsic $2\mbox{--}10$\,keV luminosity of approximately $10^{45.7}$\,erg/s and a black hole mass of $\simeq10^{9.6}\,M_\odot$ \citep[based on the single-epoch spectroscopy;][]{2022Wu}, placing it among the luminous quasars that were rapidly growing near cosmic noon. We report that its rest-frame spectrum contains two candidate redshifted absorption features. If confirmed, these features would represent the highest-redshift UFI candidate reported to date and provide a rare probe for materials around the SMBH at cosmic noon.

\section{Analysis and results}\label{sec:observations}

We jointly analyzed the 2022 \textit{XMM-Newton} (ObsID: 0913070401, duration: 68ks) and \textit{NuSTAR} (ObsID: 60861002002 \& 60861002004, total duration: 165ks) observations (PI: Zhao). Data reduction followed the standard pipeline described in Appendix A of \citet{Zhao2026}, with the SAS software (v22.1.0) and current calibration files by June 2026. We tested different background regions to confirm the stability of line features.
We fit EPIC-pn, MOS1, MOS2, FPMA, and FPMB spectra in XSPEC \citep[v12.13.1][]{1996Arnaud} using the C-statistic. The spectra were optimally binned with \texttt{ftgrouppha} \citep{2016Kaastra}. We adopted the $0.3\mbox{--}10$\,keV band for \textit{XMM-Newton} spectra and the $3\mbox{--}24$\,keV band for \textit{NuSTAR} spectra, corresponding to a combined rest-frame coverage of $\sim1\mbox{--}70$\,keV. Parameter uncertainties were estimated at the 90\% confidence level ($\Delta$C-stat=2.706).

%%%%%%%%%%%%%%%%%%%%%%%%%%%%%%%%%%%%%%%%%%%%%%%%%%
\begin{figure}[h]
    \centering        
    \includegraphics[width=\columnwidth,height=9cm]{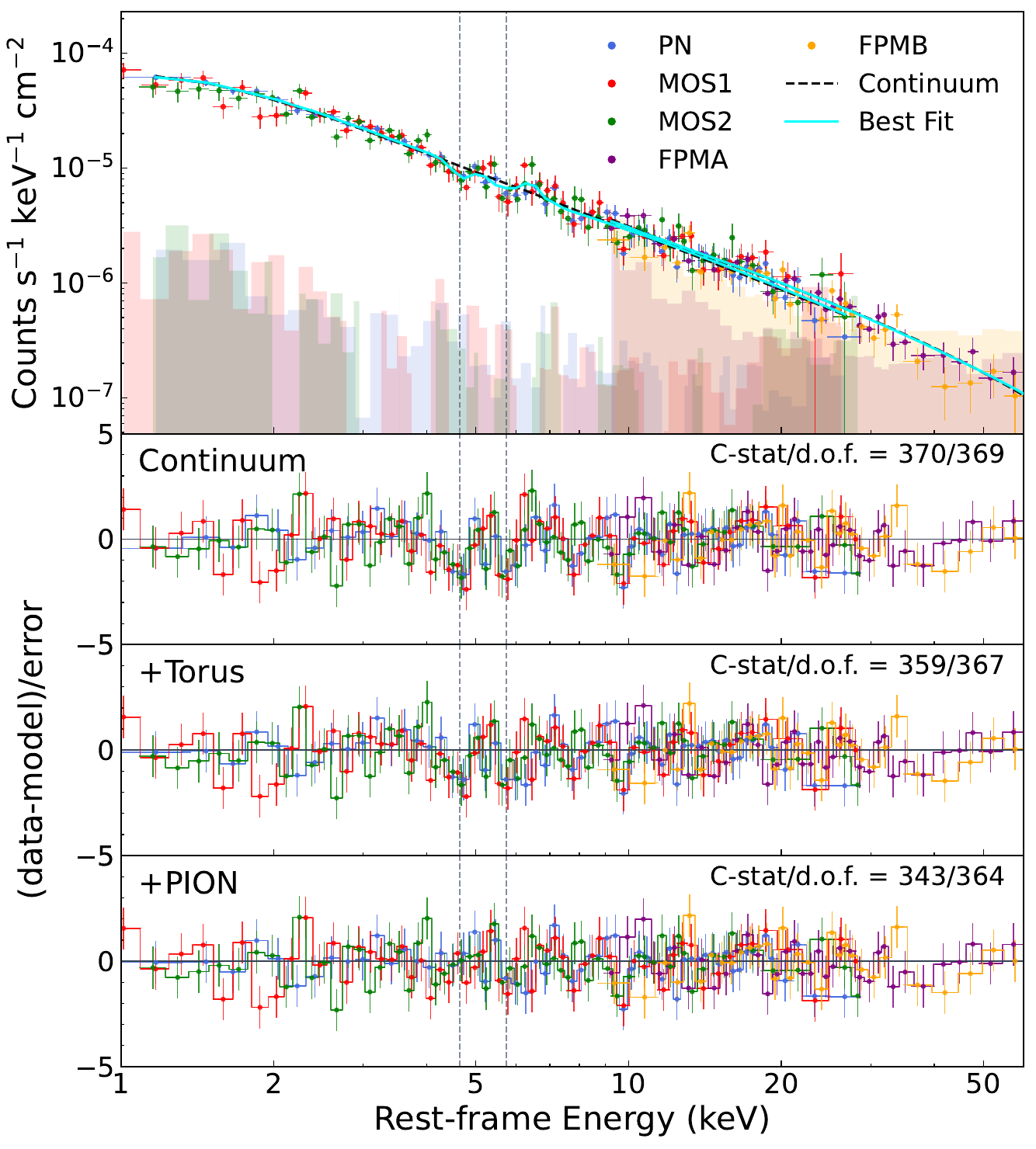}
    \caption{\textit{Top}: X-ray spectra of J0948. \textit{XMM-Newton}/EPIC-pn, MOS1, MOS2, and \textit{NuSTAR}/FPMA, FPMB data are shown in blue, red, green, purple, and orange, respectively, with corresponding backgrounds in lighter shades. The dashed black and solid cyan curves denote the continuum and best-fit models. 
    \textit{Remaining panels}: Residuals for successive models from the continuum alone to the continuum plus torus emission and one photoionized absorber. The fit statistic is reported in each panel.
    Vertical dashed lines mark absorption lines associated with the UFI.
    }
    \vspace{-2mm}
    \label{fig:fits}
\end{figure}
%%%%%%%%%%%%%%%%%%%%%%%%%%%%%%%%%%%%%%%%%%%%%%%%%%

Our baseline model (Model A) is an absorbed Comptonization model: \texttt{constant*TBabs*zTBabs*nthComp}. The multiplicative constant accounts for the differences in cross-calibration. \texttt{TBabs} and \texttt{zTBabs} describe the neutral hydrogen absorption from the Milky Way \citep[$N_\mathrm{H}^\mathrm{Gal}=10^{20}\,\mathrm{cm}^{-2}$;][]{2016HI4PI} and the host galaxy, respectively. \texttt{nthComp} models the AGN X-ray continuum, characterized by photon index $\Gamma$, seed photon temperature $kT_\mathrm{bb}=5\,\mathrm{eV}$ \citep[appropriate for a thin disk around a $M_\mathrm{BH}\sim10^{9.6}\,M_\odot$ black hole;][]{1973Shakura}, and electron temperature $kT_\mathrm{e}$. No additional soft excess component is required over the available rest-frame band. Figure \ref{fig:fits} shows the spectra and residuals, and Table \ref{app:tab:fits} lists the best-fit parameters. As shown in Appendix~\ref{app:sec:eff}, we found that potential EPIC calibration uncertainties are insufficient to fully account for the observed absorption features, favoring an astrophysical origin.

%%%%%%%%%%%%%%%%%%%%%%%%%%%%%%%%%%%%%%%%%%%%%%%%%%
\begin{figure}[h]
    \centering        
    \includegraphics[width=\columnwidth]{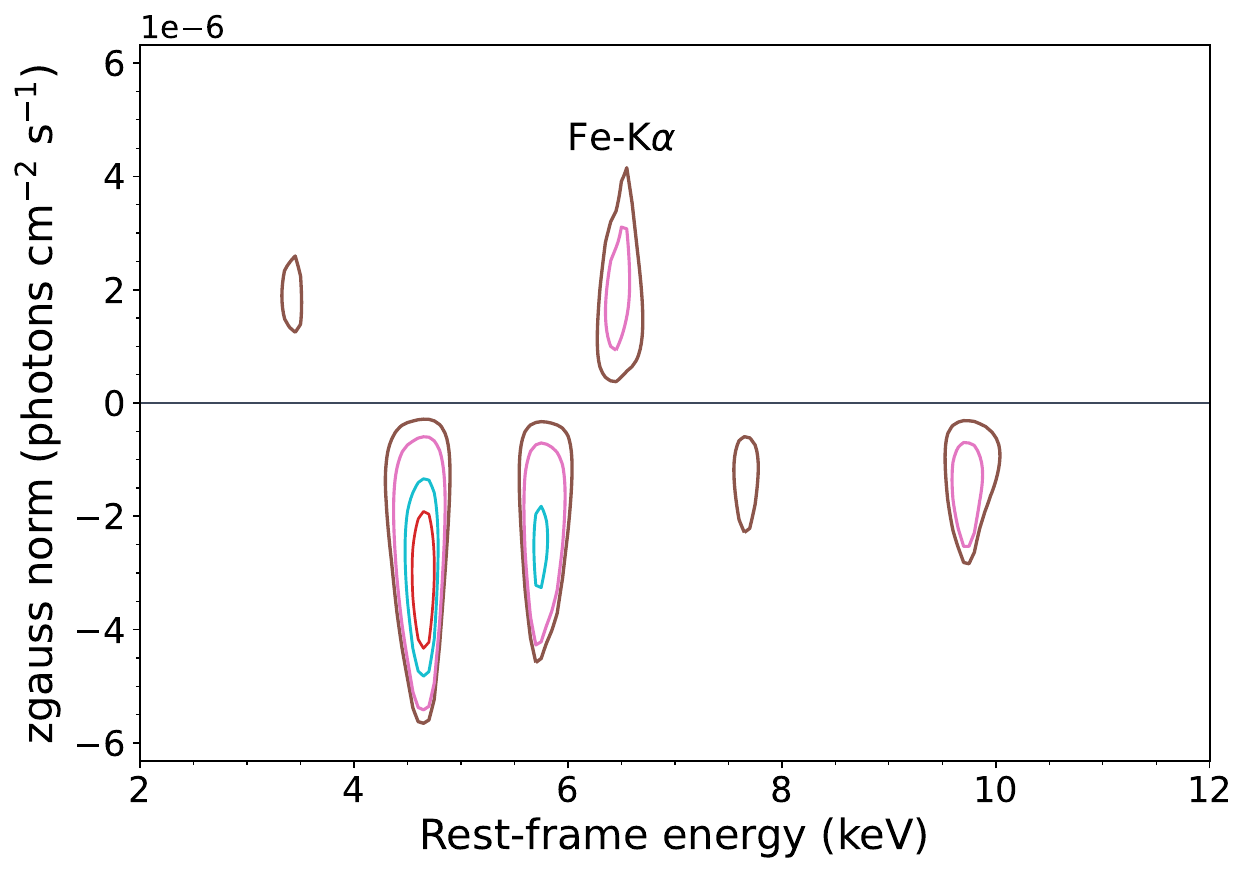}
    \caption{Blind line scan over the rest-frame 2-12\,keV band. The contours show $\Delta\mathrm{C}$ as a function of energy ($E_\mathrm{zgaus}$) and normalization ($Norm_\mathrm{zgaus}$) for a narrow Gaussian ($\sigma=10$\,eV) added to the continuum (Model A). Brown, pink, cyan, and red contours denote the 68\%, 90\%, 99\%, and 99.7\% confidence levels for two parameters ($\Delta\mathrm{C}=-2.3, -4.61, -9.21, -11.83$), respectively.
    }
    \vspace{-5mm}
    \label{fig:line-scan}
\end{figure}
%%%%%%%%%%%%%%%%%%%%%%%%%%%%%%%%%%%%%%%%%%%%%%%%%%

We searched for discrete line features by performing a blind line scan over the continuum model. Briefly, we added a \texttt{Gauss} component with a fixed line width of $\sigma=10$\,eV, sufficient for the charge-coupled device (CCD) resolution. We scanned rest-frame energies of $2\mbox{--}12$\,keV with a step of 10\,eV.
The statistical improvement ($\Delta$C-stat) was recorded at each parameter grid. The contours of 68\%, 90\%, 99\%, and 99.7\% confidence levels, assuming the Gaussian approximation, are shown in Fig. \ref{fig:line-scan}. Considering features above the $99\%$ confidence level, the scan revealed absorption lines at $\simeq4.6$ and $5.7$\,keV. A weak Fe-K$\alpha$ emission feature ($\sim$6.4\,keV) is also present at a local confidence level of 95\%. 

We modeled the Fe K$\alpha$ emission with the \texttt{borus12} model \citep{2018Balokovic}, which is a Monte Carlo (MC) radiative transfer model for the neutral scattering emission from a torus (Model B). Because the data cannot constrain its geometry, we fixed the iron abundance ($A_\mathrm{Fe}=1\,Z_\odot$), covering fraction ($C_\mathrm{F}=0.5$), and the inclination angle ($\theta=60^\circ$) while leaving the scattering column density ($N_\mathrm{H}^\mathrm{Tor}$) and normalization free. The spectral slope and coronal temperature were tied to those of \texttt{nthComp}. This component improves the fit by $\Delta$C-stat$=-11$ for two additional free parameters (i.e., a Gaussian-approximate significance of $\sim 2.9\sigma$), driven equally by the Fe-K$\alpha$ and the Compton hump.

By modeling with Gaussians, we measured two absorption lines at $4.6^{+0.1}_{-0.1}$ and $5.7^{+0.2}_{-0.1}$\,keV, with statistical improvements of $\Delta$C-stat$=-13$ and -8, respectively.
Based on Model B, we performed MC simulations that account for the look-elsewhere effect to evaluate the true significance of these lines, obtaining global confidence levels of 98.1\% ($2.4\sigma$) and 88.8\% ($1.6\sigma$; see details in Appendix~\ref{app:subsec:MC}). Their equivalent widths were measured as $-34^{+16}_{-10}$\,eV and $-39^{+17}_{-20}$\,eV, respectively, values comparable with UFIs in the literature \citep[e.g.,][]{2026Peca}.

The absorption lines were further physically modeled by including a photoionization model (Model C), \texttt{PION} \citep{2016Mehdipour}, that calculates the ionization balance of plasma for an input spectral energy distribution (SED). The SED construction is described in Appendix~\ref{app:subsec:SED}. We obtained an AGN bolometric luminosity of $L_\mathrm{bol}=9.8\times10^{46}\,\mathrm{erg/s}$ and an ionizing luminosity (1-1000\,Ryd) of $L_\mathrm{ion}=3.3\times10^{46}\,\mathrm{erg/s}$. For a black hole mass of $\sim10^{9.6}\,M_\odot$, this luminosity implies an Eddington ratio of $\lambda_\mathrm{Edd}=L_\mathrm{bol}/L_\mathrm{Edd}\simeq0.2$. Using the scripts from \citet{2019Parker}, we generated an XSPEC table model of \texttt{PION}. The model is characterized by three free parameters: the ionization parameter ($\log\xi$), column density ($N_\mathrm{H}$), and LOS redshift ($z_\mathrm{LOS}$). The unconstrained turbulent velocity $\sigma_v$ was fixed at the best-fitting value of 1000\,km/s (see top of Fig. \ref{app:fig:contours}).

Two absorption lines could be simultaneously explained by one \texttt{PION} component, identified as the Fe XXV/XXVI K$\alpha$ lines and their higher-order K$\beta$ transitions, though the 5.7\,keV feature was only partially explained by the Fe-K$\beta$ lines. We explored the best-fit solution with the \texttt{steppar} command in XSPEC. The bottom panel of Fig. \ref{app:fig:contours} shows the LOS redshift ($z_\mathrm{LOS}$) versus ionization parameter ($\log\xi$) contours. Table \ref{app:tab:fits} lists the best-fit parameters, and Fig. \ref{fig:fits} presents the residual spectrum of Model C. The component has an ionization parameter of $\log(\xi/\mathrm{erg\,cm\,s}^{-1})=4.1^{+0.6}_{-0.7}$, and an apparent column density of $N_\mathrm{H}=1.4^{+3.8}_{-0.9}\times10^{24}\,\mathrm{cm}^{-2}$. Its observer-frame LOS redshift is $z_\mathrm{LOS}=3.25^{+0.10}_{-0.07}$. If this shift is purely kinematic from the Doppler shift,\footnote{$(1+z_o)=(1+z_\mathrm{c})(1+z_a)$ and $(1+z_{a})=\sqrt{(1+\beta)/(1-\beta)}$, where $z_o$ is the observed redshift, $z_c$ is the cosmological redshift, $z_a$ is the intrinsic redshift, and $\beta=v/c$ is the gas velocity in units of $c$} the gas is moving at $v_\mathrm{in}=+0.37^{+0.02}_{-0.01}\,c$ along our LOS, where the positive value denotes an inflow. This component improves the fit by $\Delta$C-stat/d.o.f.$=-16/3$ (i.e., a local Gaussian-approximate significance of $\sim3.3\sigma$) over Model B. We also tested an additional \texttt{PION} component, but it gave only marginal improvements of $\Delta$C-stat/d.o.f.$\simeq-5/3$ by modeling the remaining residuals at $\sim5.7$\,keV or the feature at $\sim9.7$\,keV. Although they might indicate an inflow ($|v_\mathrm{in}|\sim0.15\mbox{--}0.19\,c$) and an outflow ($|v_\mathrm{out}|\sim0.32\mbox{--}0.35\,c$), given their significance, we adopted only one \texttt{PION} component in the final fit.

\section{Discussion and conclusions}\label{sec:discussion}

Our analysis revealed two rest-frame absorption lines with look-elsewhere-corrected significances of $\sim2.4~\mathrm{and}~1.6\sigma$. If identified with Fe-K transitions and attributed solely to kinematic Doppler shifts, their redshifts imply a relativistic inflow moving at $+0.37^{+0.02}_{-0.01}c$. An alternative interpretation are blueshifted transitions of lower-Z elements. Specifically, Ar XVIII Ly$\alpha$ ($3.32$\,keV) and Ca XX Ly$\alpha$ ($4.11$\,keV) at $v_\mathrm{out}\simeq-0.32c$ can simultaneously reproduce both absorption lines, while Fe XXVI Ly$\alpha$ at the same velocity can also account for the weaker residual at 9.7\,keV. This alternative provides a larger statistical improvement ($\Delta$C-stat/d.o.f.$\simeq-26/5$) than the UFI scenario. However, it requires extremely super-solar abundances of Ar ($A_\mathrm{Ar}=72^{+131}_{-44}Z_\odot$) and Ca ($A_\mathrm{Ca}=13^{+37}_{-10}Z_\odot$), which are difficult to reconcile with the expected chemical abundances, even at cosmic noon with frequent supernovae outbursts. Recent JWST observations have also reported subsolar Ar abundance in high-z galaxies \citep[e.g.,][]{2026Foley}. Therefore, we adopted the Fe-K UFI interpretation for the following discussion, but the alternative UFO scenario cannot be definitively excluded.

The shift may contain both gravitational and kinematic contributions. In the extreme case of a locally stationary absorber, pure gravitational redshift, $(1+z_a)=(1-2r_\mathrm{g}/r)^{-1/2}$, places the gas at $r=3.7^{+0.1}_{-0.1}r_\mathrm{g}$, where $r_\mathrm{g}=GM_\mathrm{BH}/c^2$ is the gravitational radius. 
For a Schwarzschild black hole (a dimensionless spin $a_\star=0$), the innermost stable circular orbit (ISCO) is $r_\mathrm{ISCO}=6r_\mathrm{g}$, while the ISCO of a maximally spinning Kerr black hole ($a_\star=0.998$) is $r_\mathrm{ISCO}=1.235r_\mathrm{g}$. The observed shift thus requires a stable prograde disk with a black hole spin of $a_\star\geq0.6$. However, transient gas in the plunging region of a Schwarzschild black hole cannot be ruled out, though this scenario requires an irradiating source within an extremely compact region and the gas motion would also contribute to the shift.

Alternatively, the shift could originate from the Doppler shift of infalling material. The velocity purely derived from this effect is $0.37^{+0.02}_{-0.01}\,c$ (Sect.\ref{sec:observations}). It corresponds to a free-fall radius of $2GM_\mathrm{H}(1-\lambda_\mathrm{Edd})/v_\mathrm{in}^2=12^{+1}_{-1}\,r_\mathrm{g}$, where $(1-\lambda_\mathrm{Edd})$ is the factor considering the reduction in effective gravity by radiation pressure \citep[$\lambda_\mathrm{Edd}\sim0.2$ for J0948;][]{2004Proga}. At this location, the impact of the gravitational redshift cannot be neglected. Therefore, we reestimated the velocity and location of the absorber considering both Doppler and gravitational shift. For radial infall in a Schwarzschild spacetime, the combined shift is
\begin{equation}
g=\frac{1}{1+z_a}=\frac{E_\mathrm{obs}}{E_0}=
\left(1-\frac{2r_\mathrm{g}}{r}\right)^{1/2}
\left(\frac{1-\beta}{1+\beta}\right)^{1/2},
\end{equation}
where $\beta\equiv v_\mathrm{in}/c$ is the local inflow speed in units of $c$. The characteristic radius was calculated by assuming that gas free falls radially from infinity, $\beta=\sqrt{2r_\mathrm{g}(1-\lambda_\mathrm{Edd})/r}$, considering the radiation pressure. Then, it gives $\beta=0.31^{+0.01}_{-0.01}$ and $r=17^{+1}_{-1}\,r_\mathrm{g}$. Although projection effects, non-radial motion, and magnetic forces are overlooked here, the estimated radius provides a characteristic location for the absorber. It gives an absorber number density of $n_\mathrm{H}=L_\mathrm{ion}/( r^2\xi)=3^{+11}_{-2}\times10^{10}\,\mathrm{cm}^{-3}$ and a thickness of $\Delta r=N_\mathrm{H}/n_\mathrm{H}=5.2^{+28.1}_{-4.7}\times10^{13}\,\mathrm{cm}$. The implied filling factor $f_\mathrm{V}\approx\Delta r/r\sim5.4^{+28.8}_{-4.9}\times10^{-3}$ corresponds to a clumpy structure, similar to UFOs in other high-Eddington AGNs \citep[e.g.,][]{2025Xrism,2025Xu,2026Mizumoto}.

The mass inflow rate was estimated as
\begin{equation}
\dot{M}_\mathrm{in}\simeq4\pi\mu m_\mathrm{p}n_\mathrm{H}r^2v_\mathrm{in}f_\mathrm{\Omega}f_\mathrm{V}
=26^{+70}_{-17}\,M_\odot\,\mathrm{yr}^{-1}
\end{equation}
using $N_\mathrm{H}=n_\mathrm{H}f_\mathrm{V}r$, where $f_\mathrm{\Omega}=\Omega/4\pi$ ($\Omega$: solid angle) is the opening fraction assumed at a fiducial value of 0.5 \citep[e.g.,][]{2018Nardini}, $\mu=1.23$ is the mean atomic weight, and $m_\mathrm{p}$ is the proton mass.  
For comparison, the mass accretion rate derived from the bolometric luminosity is $\dot{M}_\mathrm{acc}=L_\mathrm{bol}/(\eta c^2)
\simeq18\,M_\odot\,\mathrm{yr}^{-1}$ \citep[a fiducial accretion efficiency $\eta=0.1$;][]{1973Shakura}. 
% The candidate inflow therefore 
The comparable rates suggest a potentially substantial inward mass flux, although whether this material ultimately contributes to SMBH growth remains uncertain.
% might represent a substantial mass inflow flux for the AGN accretion}. 
Alternatively, requiring that the inflow rate does not exceed the accretion rate, $\dot{M}_\mathrm{in}\lesssim \dot{M}_\mathrm{acc}$, also provides an independent upper limit on the absorber location of $r\lesssim r_\mathrm{max}=8^{+14}_{-5}r_\mathrm{g}$, which is consistent with the inferred characteristic radius within the uncertainties. The corresponding kinetic power is $\dot{E}_\mathrm{K}=0.5\dot{M}_\mathrm{in}v_\mathrm{in}^2=7^{+19}_{-5}\times10^{46}\,\mathrm{erg/s}$, comparable to the bolometric luminosity. Applying the special-relativistic correction to the column density for a radial inflow yielded
% If we consider the special relativity effect and assume a radial inflow, the corrected column density is 
$N_\mathrm{H;cor}=7^{+17}_{-4}\times10^{23}\,\mathrm{cm}^{-2}$\footnote{$N_\mathrm{H;cor}=N_\mathrm{H}(1+\beta\mathrm{cos}\theta)^2/(1-\beta^2)$, where $\beta=v_\mathrm{wind}/c$ and $\theta$ is the angle between incident photons and gas velocity ($\theta=180^{\circ}$ for radial inflows).} \citep{2020Luminari}. The resulting mass inflow rate and kinetic power are $\dot{M}_\mathrm{in;cor}\simeq13^{+31}_{-7}\,M_\odot\mathrm{yr}^{-1}$ and $\dot{E}_\mathrm{K;cor}=4^{+9}_{-2}\times10^{46}\,\mathrm{erg/s}$, respectively. We note that these are order-of-magnitude estimates, and the general relativity effects on irradiation fields (e.g., gravitational redshift and light-bending effect) have not been incorporated into the modeling. Quantifying these effects requires additional assumptions about the geometry that are beyond the scope of this work.

On large scales, chaotic cold accretion may supply clumpy materials from the kiloparsec to subparsec scales to fuel AGNs \citep[e.g.,][]{Gaspari2013}. Closer to the SMBH, several mechanisms could give rise to UFIs. One possibility is the fragmentation of inner accretion flows. The fragmentation could originate from (1) instability of a misaligned accretion disk that undergoes tearing \citep[e.g.,][]{Dogan2018} or (2) magnetically channeled, non-equatorial accretion flows \citep{Fukumura2026}. In both scenarios, UFIs are expected to recur episodically, and no UFO detection is required. A possible discriminant between these scenarios is their predicted velocity structure. In the disk-tearing scenario, distinct multi-velocity inflow components are predicted, as the collision between rings can transport materials to different circularization radii, whereas the magnetically channeling explanation expects a coherent velocity field of clumps. The possible secondary inflow from the remaining 5.7\,keV residuals would be qualitatively consistent with multiple streams, and if confirmed, its distinct velocity could support the disk-tearing interpretation, but its low significance precludes any firm conclusions. Likewise, the weak reflection fraction $R\equiv L_\mathrm{2-10}^{torus}/L_\mathrm{2-10}^{corona}\simeq0.17$ does not support a warped inner disk. 

The absorber could instead result from the fallback of ejecta in the forms of ``aborted jets" or ``failed winds" \citep[e.g.,][]{2004Proga,Ghisellini2004} that were initially launched outward but failed to escape the gravitational potential well of the central SMBH. In an aborted jet picture, plasma blobs, which are intermittently launched along the spin axis, fail to escape and may shock against subsequent ejecta, producing hard X-ray emission. Interestingly, the inferred UFI kinetic power is similar to the hard X-ray emission of J0948 ($\sim10^{46}\,\mathrm{erg/s}$), and J0948 is also detected at 1.4 GHz with $F_\mathrm{1.4GHz}=3.1\pm0.2\,\mathrm{mJy}$ (from the VLA FIRST survey), leaving weak jet activity as a possible but unconstrained interpretation. However, the hard X-ray light curve is stable during the observation, and the inclination is unconstrained, so the expected association with coronal flares and a polar flow cannot be tested.
Another similar scenario is the failed wind, where the absorbing material is part of an ongoing wind that becomes overionized by the intense continuum radiation. As the wind loses radiative support, it stalls and eventually falls back toward the central engine. Such a configuration naturally allows coexisting fallback and outflowing material, potentially indicated by the weak 9.7\,keV feature. Deep exposures are required to verify this tentative feature and uncover the underlying mechanism.

%%%%%%%%%%%%%%%%%%%%%%%%%%%%%%%%%%%%%%%%%%%%%%%%%%
\begin{figure}[h]
    \centering        
    \includegraphics[width=\columnwidth]{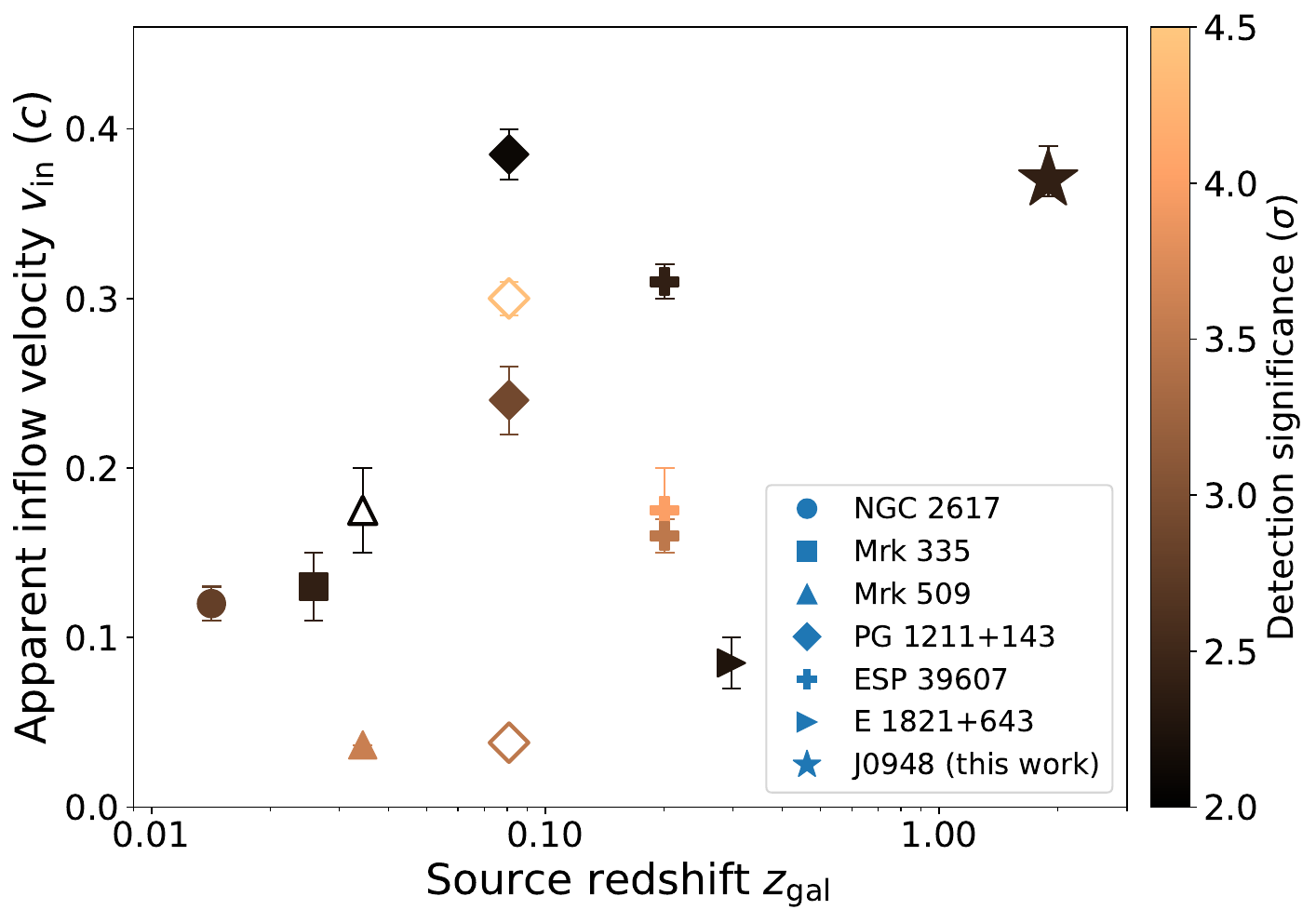}
    \caption{Comparison of reported UFI candidates in the plane of the apparent LOS inflow velocity ($v_\mathrm{in}$) and the target redshift ($z_\mathrm{gal}$). Colors indicate the reported detection significance from MC simulations, while the hollow symbols represent the Gaussian-approximate significances.
    }
    \vspace{-3mm}
    \label{fig:ufi-comparison}
\end{figure}
%%%%%%%%%%%%%%%%%%%%%%%%%%%%%%%%%%%%%%%%%%%%%%%%%%

Regardless of its origin, the large redshift points to an absorber from a compact region within tens of $r_\mathrm{g}$. Compared with previously reported UFI candidates (Fig. \ref{fig:ufi-comparison}), the UFI candidate in J0948 occupies the high-velocity end of the observed distribution and represents the highest-redshift UFI candidate reported to date. Its apparent pure-Doppler velocity of $0.37^{+0.02}_{-0.01}c$ is comparable to the fastest events in PG 1211+143 and exceeds most other candidates. J0948 therefore extends this extreme inflow phenomenon from nearby AGNs to a luminous quasar at cosmic noon, providing a rare probe for gas dynamics around SMBHs at the peak epoch of AGN activity.
Its clumpy structure is consistent with an inhomogeneous, possibly episodic mode of accretion rather than a smooth inflow when the black hole grows rapidly \citep{2026Xu}.

Unlike the multi-ion UFI in PG 1211+143 \citep{2018Pounds} or the relatively slower absorber resolved by XRISM in Mrk 509 \citep{Dadina2026}, the UFI interpretation in J0948 is still driven primarily by unresolved Fe-K troughs of modest significance. Robust identification requires repeated detections or multiple transitions sharing a common velocity shift. At present, the interpretation is limited by significances, ionic and redshift degeneracies at CCD resolution, and the uncertain geometry of absorbers. Deeper \textit{XMM-Newton} observations are necessary to confirm the detection of these absorption lines, while future high-resolution spectra from NewATHENA will be essential to breaking the degeneracy among ionic identifications and precisely measuring their widths.

\begin{acknowledgements}
YX and MM acknowledge support from the MCIU program Unidad de Excelencia Mar\'a de Maeztu CEX2020-001058-M and the MaX-CSIC Excellence Award MaX4-SOMMA-ICE. XZ acknowledges the support from NASA funding 80NSSC24K1031. VEG acknowledges funding under NASA contract 80NSSC24K1403. CP is supported by INAF Large Grant 2023 BLOSSOM O.F. 1.05.23.01.13. 
\end{acknowledgements}
\vspace{-9mm}
\bibliographystyle{aa}
\bibliography{ref.bib}

\begin{appendix}
\nolinenumbers
\section{Calibration effects}\label{app:sec:eff}
\begin{figure}[h]
    \centering        
    \includegraphics[width=\columnwidth]{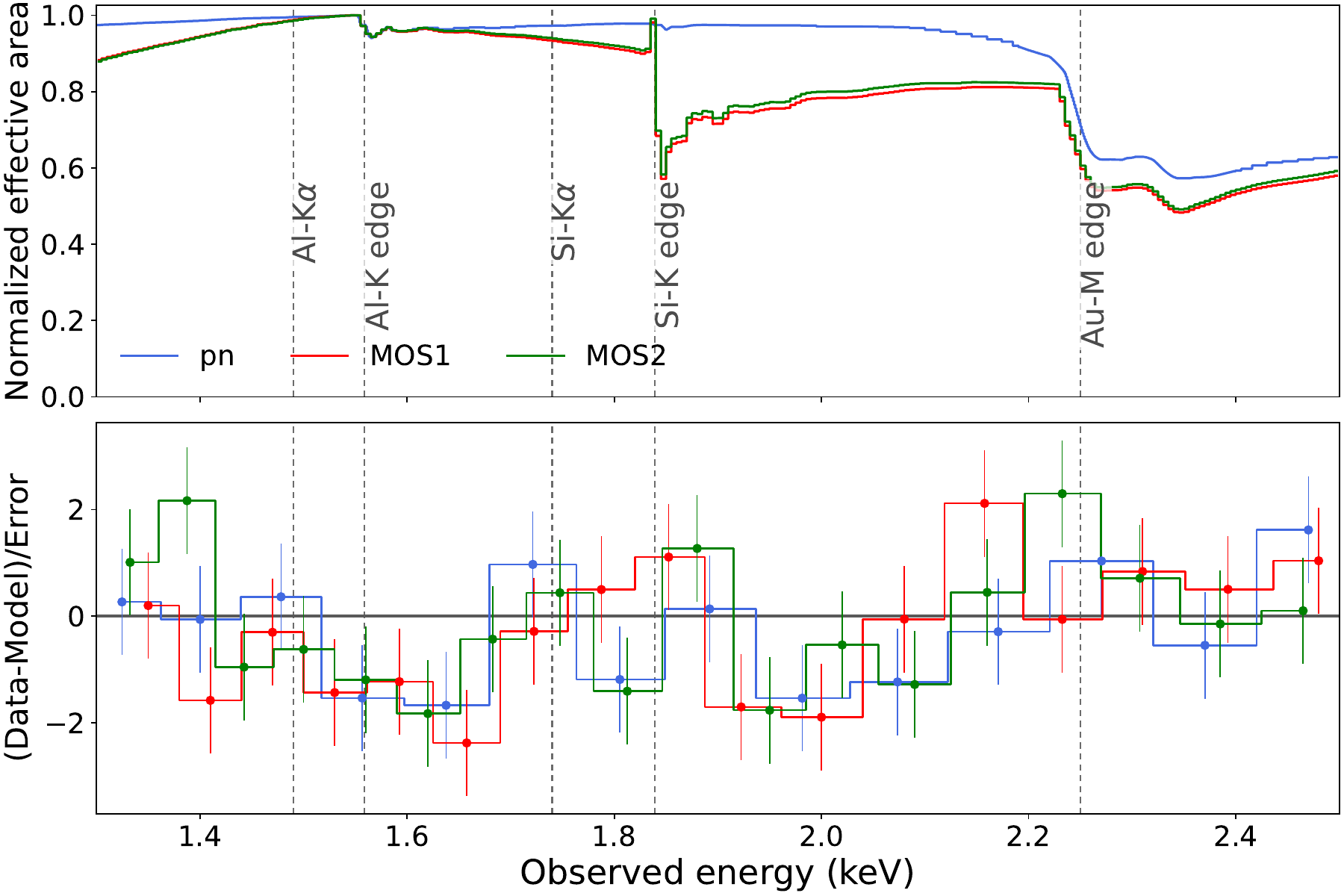}
    \caption{Zoom-in plots of the EPIC normalized effective area (\textit{top}) and the residuals against the continuum (\textit{bottom}) with the Al-K, Si-K, and Au-M edges, and Al-K$\alpha$ and Si-K$\alpha$ instrumental lines marked by vertical dashed lines. 
    }
    \vspace{-2mm}
    \label{app:fig:eff}
\end{figure}

In the observer frame, the two absorption features fall in an energy range containing several sharp instrumental edges in the EPIC effective area, including the Al K ($\sim$1.56\,keV), Si K ($\sim$1.84\,keV), and Au-M ($\sim2.25$\,keV) edges, and Al-K$\alpha$ ($\sim1.49$\,keV) and Si-K$\alpha$ ($\sim1.74$\,keV) lines, where calibration residuals are known to occur in \textit{XMM-Newton} spectra. Such residuals may therefore introduce systematic uncertainties in this energy band. To assess their possible impact, we zoom in on this region in Fig. \ref{app:fig:eff}, showing the normalized EPIC effective areas together with the data-to-continuum residuals, with the instrumental edges marked by vertical dashed lines.

% %%%%%%%%%%%%%%%%%%%%%%%%%%%%%%%%%%%%%%%%%%%%%%%%%%
% %%%%%%%%%%%%%%%%%%%%%%%%%%%%%%%%%%%%%%%%%%%%%%%%%%

Both absorption features are visible in all three EPIC detectors. The $\sim2$\,keV feature is well free from edges or lines, whereas the $\sim1.6$\,keV feature partially overlaps the Al-K edge. We estimate that the latter removes a fraction of $|F_{\rm line}|/F_{\rm cont} = 12^{+6}_{-6}\%$ of the continuum flux in the $1.4\mbox{--}1.7$\,keV band. For comparison, systematic discrepancies associated with imperfect EPIC calibration are typically $\lesssim5\%$\footnote{xmmweb.esac.esa.int/docs/documents/CAL-SRN-0410-1-1.pdf}. Thus, although contributions from calibration uncertainties around the Al-K edge cannot be excluded, their expected amplitude is insufficient to account for the full observed feature. Together with the consistent residuals seen across the three EPIC detectors, this favors an astrophysical origin for the absorption features.

\section{Spectral analysis}\label{app:sec:fits}
\begin{figure}[h]
    \centering        
    \includegraphics[width=\columnwidth]{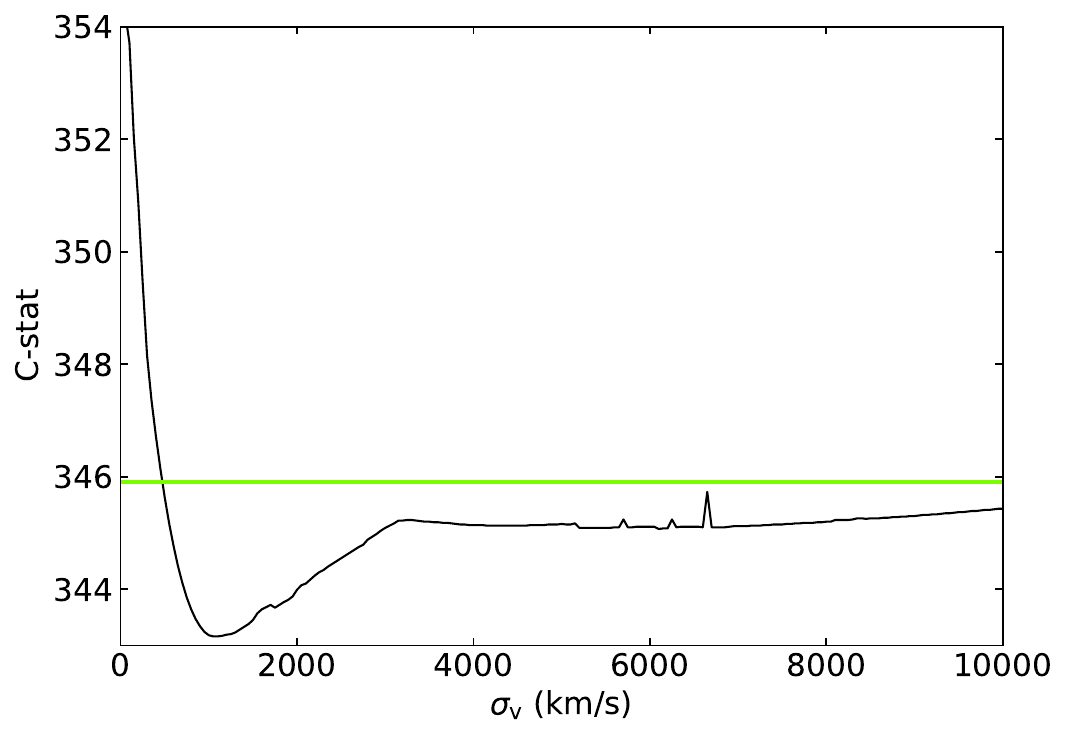}
    \includegraphics[width=\columnwidth]{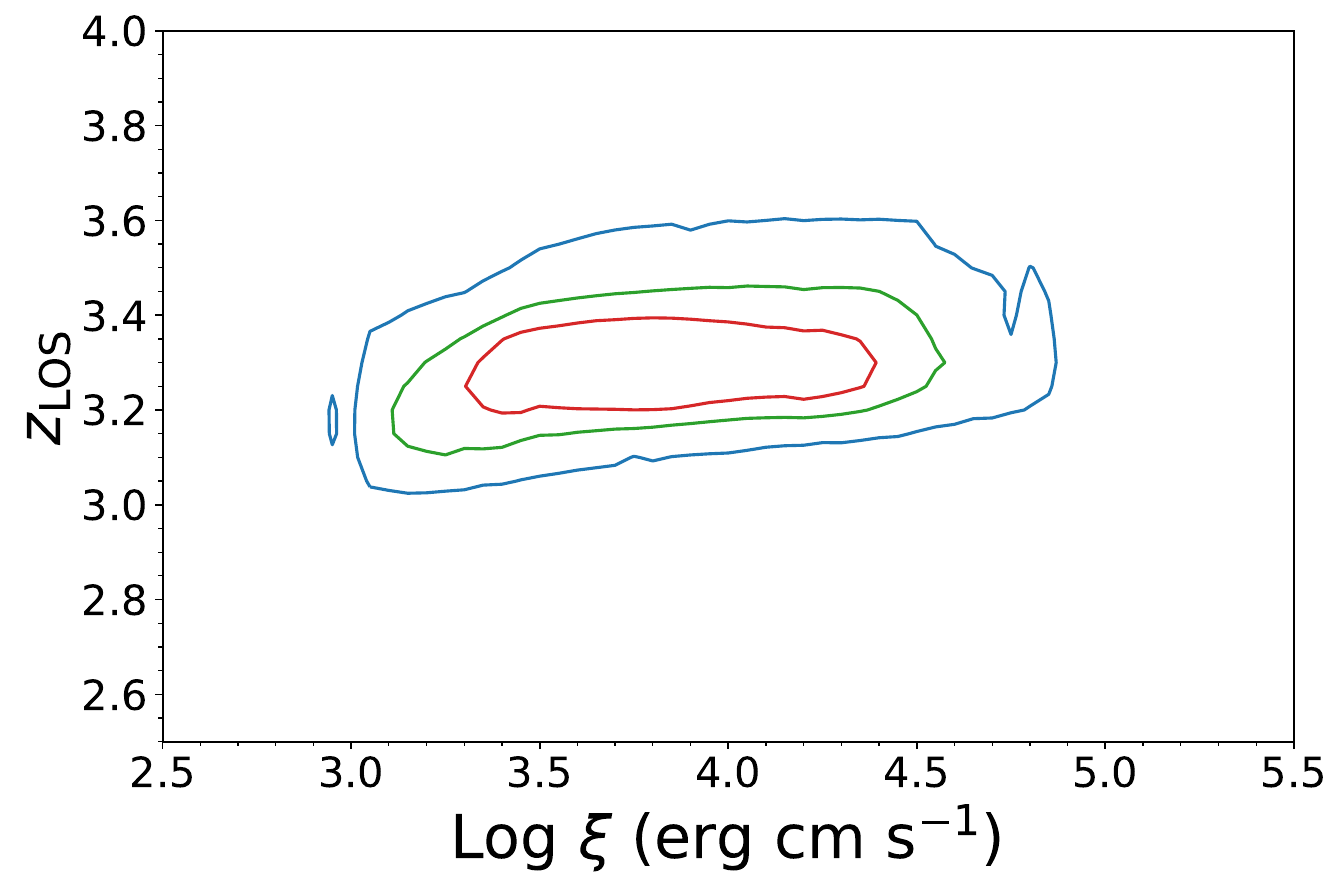}
    \caption{\textit{Top}: One-dimensional contour plot of the turbulent velocity of the \texttt{PION} component if it is free, where the horizontal line marks the 90\% confidence level ($\Delta$C-stat=2.706) for one parameter.
    \textit{Bottom}: Confidence contours of the LOS redshift $z_\mathrm{LOS}$ vs. ionization parameter $\log\xi$ for the \texttt{PION} component in Model C. Red, green, and blue contours show the 68, 90, and 99\% ($\Delta$C-stat=2.3, 4.6, 9.2) contour levels with respect to the minimum, respectively.
    }
    \vspace{-6mm}
    \label{app:fig:contours}
\end{figure}

This appendix presents the details of best fits of various models (Table \ref{app:tab:fits}), contour plots of parameters in the photoionization component (Fig. \ref{app:fig:contours}), MC simulations for evaluating line significance, and the SED construction for photoionization modeling.

% %%%%%%%%%%%%%%%%%%%%%%%%%%%%%%%%%%%%%%%%%%%%%%%%%%
% %%%%%%%%%%%%%%%%%%%%%%%%%%%%%%%%%%%%%%%%%%%%%%%%%%

\begin{table}
\setlength{\tabcolsep}{2pt}
\renewcommand{\arraystretch}{1.3}
\centering
\caption{Best-fit parameters of different model combinations for the joint \textit{XMM-Newton} and \textit{NuSTAR} spectra.}
\begin{tabular}{lcccc}
\hline
\hline
Component & Parameter & Model A$^a$ & Model B$^b$ & Model C$^c$  \\
\hline
{\tt TBabs} &    $N^\mathrm{Gal}_\mathrm{H}$ ($10^{20}$ cm$^{-2}$)      & \multicolumn{3}{c}{$1^{\star}$} \\
\hline
{\tt zTBabs} &    $N^\mathrm{host}_\mathrm{H}$ ($10^{21}$ cm$^{-2}$)         & $2.5^{+1.1}_{-1.1}$ & $3.9^{+1.5}_{-1.4}$ & $3.8^{+1.6}_{-1.9}$  \\
\hline
\multirow{4}{1em}{\tt nthComp} & $\Gamma$ & $1.75^{+0.03}_{-0.03}$ & $1.88^{+0.09}_{-0.08}$ & $1.88^{+0.13}_{-0.08}$  \\
& kT$_{bb}$ (eV) & \multicolumn{3}{c}{$5^\star$} \\
& kT$_{e}$ (keV) & $21^{+75}_{-6}$ & $>16$ & $>15$  \\
& Norm$_\mathrm{HC}$ ($10^{-5}$) & $7.1^{+0.3}_{-0.3}$ & $7.2^{+0.3}_{-0.3}$ & $21^{+102}_{-13}$ \\
\hline
 & log($N_\mathrm{H}^\mathrm{Tor}/\mathrm{cm}^{-2}$) & - & $24.1^{+0.3}_{-0.3}$ & $<24.1$   \\
 & $A_\mathrm{Fe}$ ($Z_\odot$) & -& \multicolumn{2}{c}{$1^\star$} \\
 \texttt{borus12} & $\cos\theta$ & -& \multicolumn{2}{c}{$0.5^\star$}  \\
&     $C_F$ &  -  &  \multicolumn{2}{c}{$0.5^\star$}  \\
&  Norm$_\mathrm{Tor}$ ($10^{-3}$) & -& $0.7^{+0.6}_{-0.4}$ & $3^{+36}_{-2}$  \\
\hline
\multirow{4}{4em}{\texttt{PION}} & log($\xi$/erg cm s$^{-1}$)    & - & - & $4.1^{+0.6}_{-0.7}$  \\
 &   $N_\mathrm{H}$ ($10^{24}\,\mathrm{cm}^{-2}$)    & - & - & $1.4^{+3.8}_{-0.9}$  \\
 &    $\sigma_{v}$ (km/s)    & - & - & $1000^{\star}$  \\
  &     $z_\mathrm{LOS}$  & - & - & $3.25^{+0.10}_{-0.07}$ \\
\hline
Fit-statistics &   C-stat/d.o.f.$^\dagger$  &   370/369 &359/367  & 343/364   \\
\hline
\hline
\end{tabular}
\label{app:tab:fits}
\begin{flushleft}
{$^a$ Model A: \texttt{constant*TBabs*zTBabs*nthComp}}\\
{$^b$ Model B: \texttt{constant*TBabs*zTBabs*(nthComp+borus12)}}\\
{$^c$ Model C: \texttt{constant*TBabs*zTBabs*PION*(nthComp+borus12)}}\\
{$^{\star}$ The parameter is fixed.}\\
{$^{\dagger}$ d.o.f.: Degrees of freedom.}\\
\end{flushleft}
\vspace{-4mm}
\end{table}

\subsection{Monte Carlo simulations}\label{app:subsec:MC}
\begin{figure}[!htbp]
    \centering        
    \includegraphics[width=\columnwidth]{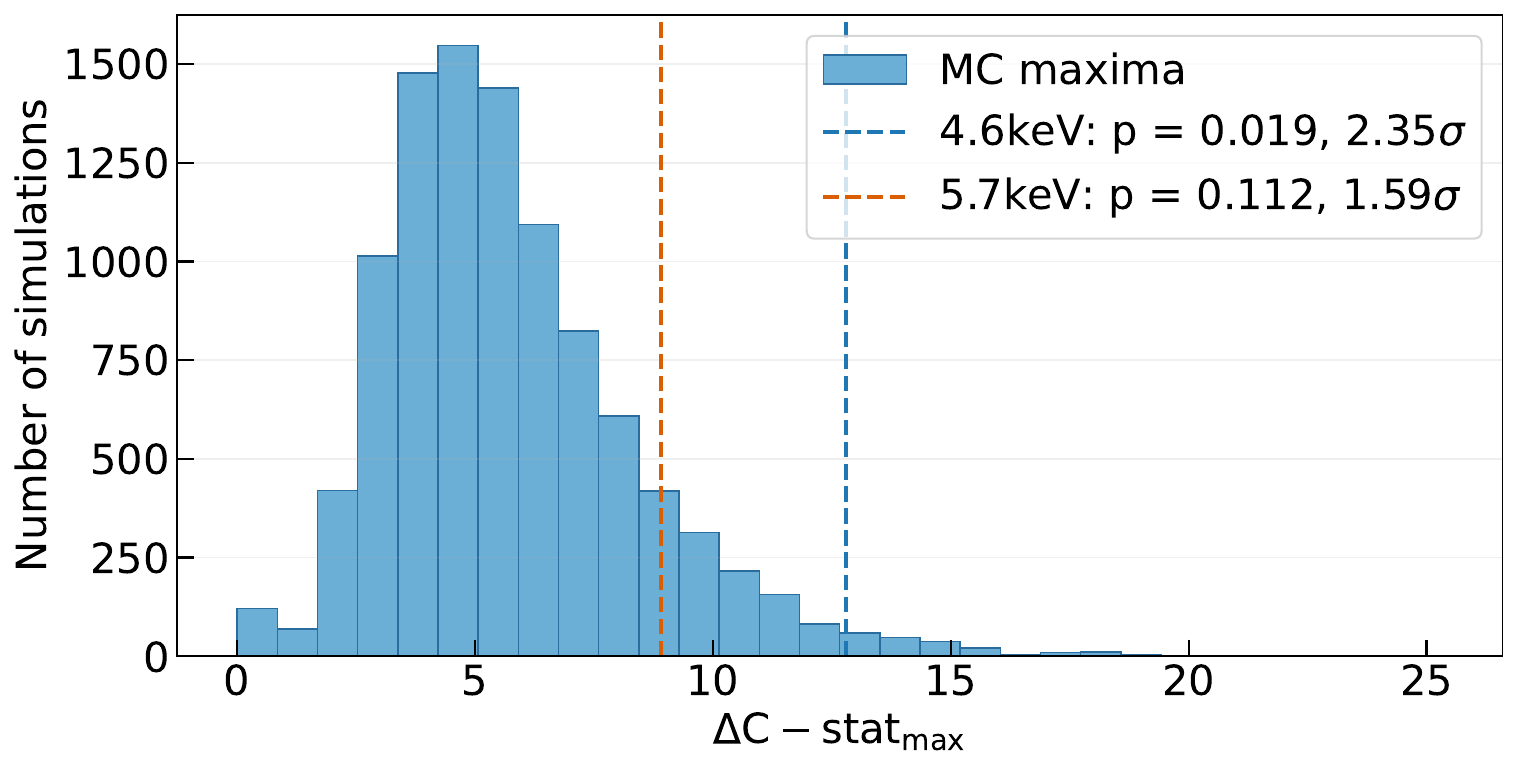}
    \caption{Distributions of the maximal $\Delta C$ from the absorption line in 10000 MC simulated spectra based on the continuum model (Model B). Vertical lines mark the observed values for two absorption features.
    }
    \vspace{-2mm}
    \label{app:fig:MC}
\end{figure}

The significance of two absorption lines is estimated through MC simulations considering the look-elsewhere effect. We performed 10000 MC simulations with the same exposures based on the continuum model (Model B) using the \texttt{fakeit} tool. For each set of five simulated spectra, we refit them with Model A and repeated the blind line scan over the best-fit continuum model, recording the maximal $\Delta$C-stat improvement of any absorption line in the simulated spectra. Figure \ref{app:fig:MC} compares this distribution with two observed features, where vertical lines mark their $\Delta$C-stat. The $p$-values of 4.6 and 5.7\,keV lines are 0.019 and 0.112, respectively, with corresponding significances of $2.4~\mathrm{and}~1.6\sigma$.

%%%%%%%%%%%%%%%%%%%%%%%%%%%%%%%%%%%%%%%%%%%%%%%%%%
%%%%%%%%%%%%%%%%%%%%%%%%%%%%%%%%%%%%%%%%%%%%%%%%%%

\subsection{Spectral energy distribution}\label{app:subsec:SED}

\begin{figure}[!htbp]
    \centering        
    \includegraphics[width=\columnwidth]{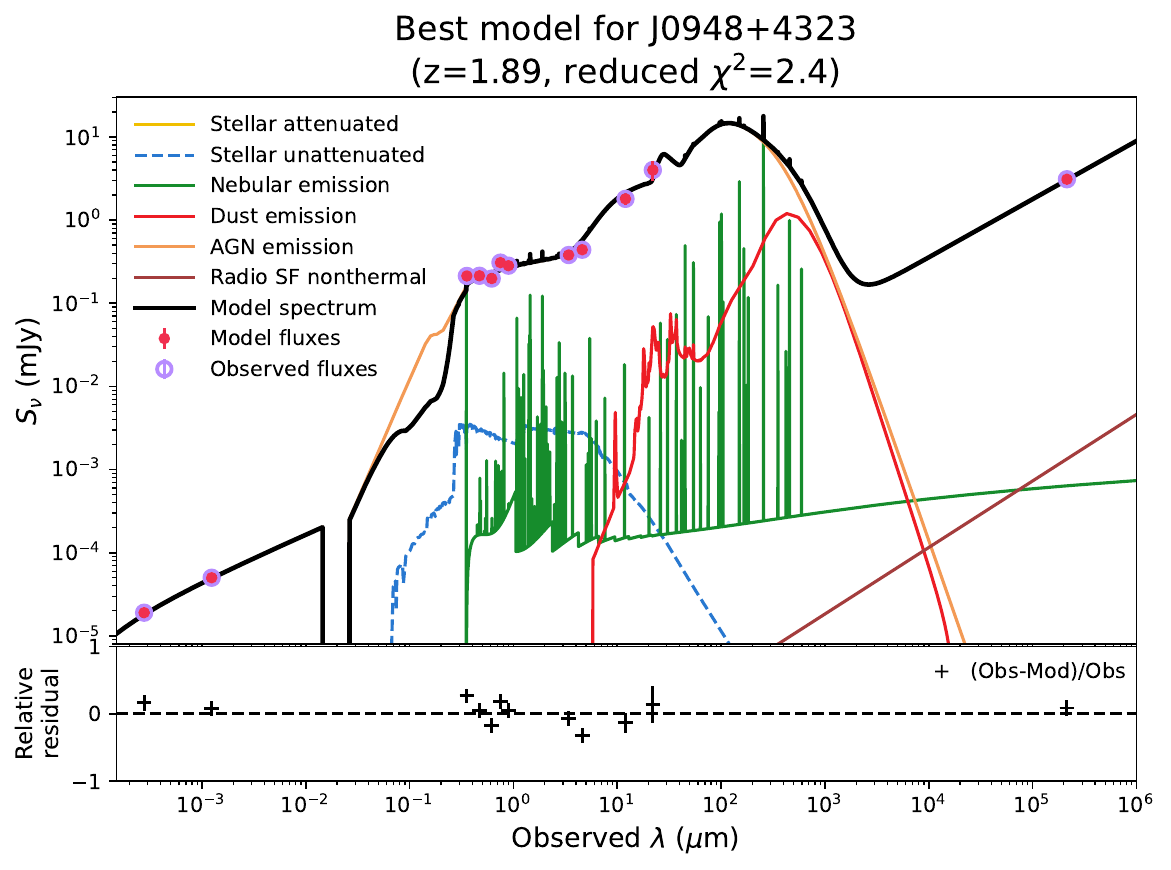}
    \caption{Best-fit broadband SED model (\textit{top}) and corresponding residuals (\textit{bottom}) with \texttt{CIGALE}. Curves represent the contributions from various components: attenuated stellar emission (yellow), unattenuated stellar emission (blue), nebular emission (green), dust emission (red), AGN emission (orange), and nonthermal radio emission (dark red). The black curve is the total model spectrum. The modeled and observed fluxes are overlaid as red and violet points, respectively.
    }
    \vspace{-2mm}
    \label{app:fig:SED}
\end{figure}

We constructed the intrinsic SED of J0948 following the PACHA procedure \citep{Zhao2026}. We performed the SED fitting with \texttt{CIGALE} \citep{2019Boquien}. The data comprise unabsorbed 0.5$\mbox{--}$2 and 2$\mbox{--}$10\,keV fluxes, dereddend SDSS DR18 photometry ($u$, $g$, $r$, $i$, and $z$ filters), WISE W1-W4 measurements, and the 1.4-GHz flux density from the VLA FIRST survey. Following the PACHA analysis, an additional 10\% uncertainty was assigned to the photometric measurements to account for cross-survey calibration differences and source variability. The intrinsic X-ray fluxes were derived from the continuum model after removing the \texttt{PION} absorption, initially modeled using a precalculated table for another AGN. Although this preliminary table does not provide a reliable ionization parameter for J0948, it allows us to estimate the intrinsic X-ray continuum for SED construction. The resulting SED was then adopted for the final photoionization modeling to ensure consistency.
The \texttt{CIGALE} model included stellar, nebular, dust-attenuation, dust-emission, AGN, X-ray, and radio components. In particular, the AGN emission was modeled with SKIRTOR, with the X-ray continuum linked to the intrinsic accretion-disk emission at 2500~\AA. The best fit is AGN dominated, with an AGN fraction of 0.99 (Fig. \ref{app:fig:SED}).

%%%%%%%%%%%%%%%%%%%%%%%%%%%%%%%%%%%%%%%%%%%%%%%%%%
%%%%%%%%%%%%%%%%%%%%%%%%%%%%%%%%%%%%%%%%%%%%%%%%%%

For the photoionization calculations, we retained the intrinsic AGN component: unattenuated UV-optical disc emission plus the unabsorbed X-ray continuum (orange curve in Fig. \ref{app:fig:SED}). The AGN bolometric luminosity was integrated in the rest frame between $1\,\mu$m and 100\,keV, corresponding to approximately 1.24\,eV-100\,keV, producing $L_\mathrm{bol}\simeq9.8\times10^{46}\,\mathrm{erg/s}$, consistent with the reported value \citep{Zhao2026}. The intrinsic ionizing luminosity (1-1000\,Ryd, i.e., 13.6eV-13.6keV) is $L_\mathrm{ion}=3.3\times10^{46}\,\mathrm{erg/s}$.

\end{appendix}

\end{document}